\documentclass[10pt,letterpaper,aps,prl,twocolumn,reprint,superscriptaddress,floatfix,footinbib]{revtex4-1}
\usepackage[margin=1in,footskip=0.25in]{geometry}
 
\usepackage{graphicx}
\usepackage{dcolumn}
\usepackage{bm}
\usepackage{hyperref}

\usepackage{graphicx,epsfig}
\usepackage{color}
\usepackage[usenames,dvipsnames]{xcolor}
\usepackage{amsmath,bbm,amssymb, amsthm}
\usepackage{dsfont} 
\usepackage{stmaryrd}
\usepackage{numprint}

\def\bb{\begin{eqnarray}}
\def\ee{\end{eqnarray}}
\newcommand{\ket}[1]{| #1 \rangle}

\def\kB{k_\text{B}}

\newcommand{\la}{\langle}
\newcommand{\ra}{\rangle}
\newcommand{\be}{\begin{equation}}
\newcommand{\eeq}{\end{equation}}

\newcommand{\Ai}{\operatorname{Ai}}

\begin{document}

\title{Efficient Quantum Measurement  Engines}

\author{Cyril Elouard}
\email{celouard@ur.rochester.edu}
\affiliation{Department of Physics and Astronomy, University of Rochester, Rochester, NY 14627, USA}

\author{Andrew N. Jordan}
\affiliation{Department of Physics and Astronomy, University of Rochester, Rochester, NY 14627, USA}
\affiliation{Center for Coherence and Quantum Optics, University of Rochester, Rochester, NY 14627, USA}
\affiliation{Institute for Quantum Studies, Chapman University, Orange, CA 92866, USA}

\date{\today}

\begin{abstract}
 We propose quantum engines powered entirely by the quantum measurement process. Our theoretical construction of the engine requires no work from the system Hamiltonian, and takes energy only from the process of observation to move a particle against a force. We present results for the work done and the efficiency for different values of the engine parameters.  Feedback is required for optimal performance. We find that unit efficiency can be approached when one measurement outcome prepares the initial state of the next engine cycle, while the other outcomes leave the original state nearly unchanged.
\end{abstract}

\maketitle


Discussions of quantum measurement typically present two aspects:  The probability to find different possible results of the measuring device, and how that result changes the quantum state of the system being interrogated.
This process is characterized as simply observing the system.  However, researchers have begun to realize that there are resources required to make measurements, and even forbid certain measurements from taking place that violate conserved quantities, such as energy \cite{PhysRevLett.112.140502}.  Measurement places energy costs on quantum information tasks, but many also allow energy extraction from the system \cite{Abdelkhalek16}. Existing results in the literature typically focus on the Wigner, Araki, Yanase (WAY) theorem \cite{wigner1952messung,araki1960measurement,yanase1961optimal} and its generalizations \cite{PhysRevLett.106.110406,ahmadi2013wigner}. One consequence of the WAY theorem is that it is impossible to have a repeatable and accurate measurements of the system's energy if the total Hamiltonian of the meter and system does not commute with the system Hamiltonian.

More recent studies of the thermodynamic aspects of measurement \citep{elouard2017role}, suggested the following principle: measurement can also bee seen as a {\it thermodynamic resource}, analogous to heat or work reservoirs, such as a battery, in classical thermodynamics. A recent work \citep{elouard2017extracting} proposed a quantum Maxwell demon being able to extract energy from measurement-induced coherences in a qubit, using Rabi oscillations to put energy into a coherent optical tone. However, as the dispersive measurement of the qubit also involved an optical field, the net result is simply using the qubit as an energy transducer from one optical mode to another. Ref.~\cite{PhysRevE.96.022108} exploits measurement to let the system do work on a classical magnetic field or a time-varying external potential. Ref. \cite{Gelbwaser13} also proposes a measurement-induced work extraction relying on non-Markovian effects in a zero temperature thermal reservoir.

In the current paper, we posit this principle by proposing measurement-fueled engines able to drive a single-particle current against a potential barrier. These engines thus do useful work such as raising an elevator, or charging a battery.
The energy comes entirely from the process of observation: Measuring the system in a basis that does not commute with its Hamiltonian allows energy to be taken away from the measurement apparatus and given to the system in such a way as to be turned into useful work. This energy transfer is stochastic in nature, so has some similarities to heat in a stochastic thermodynamics context \cite{elouard2017role}.
We stress, however, this similarity is only superficial, in that we show the existing thermodynamic bounds do not apply, and we are able to design engines with this ``quantum heat'' that can approach unit efficiency \cite{elouard2017extracting}. As recently stated, such an engine would not operate if the measuring apparatus was isolated \cite{Mohammady17}: input power must be provided to the apparatus to perform such measurements, which is taken into account for the engine's efficiency.
\begin{figure}[th]\label{Fig1}
\includegraphics[width = 0.48\textwidth]{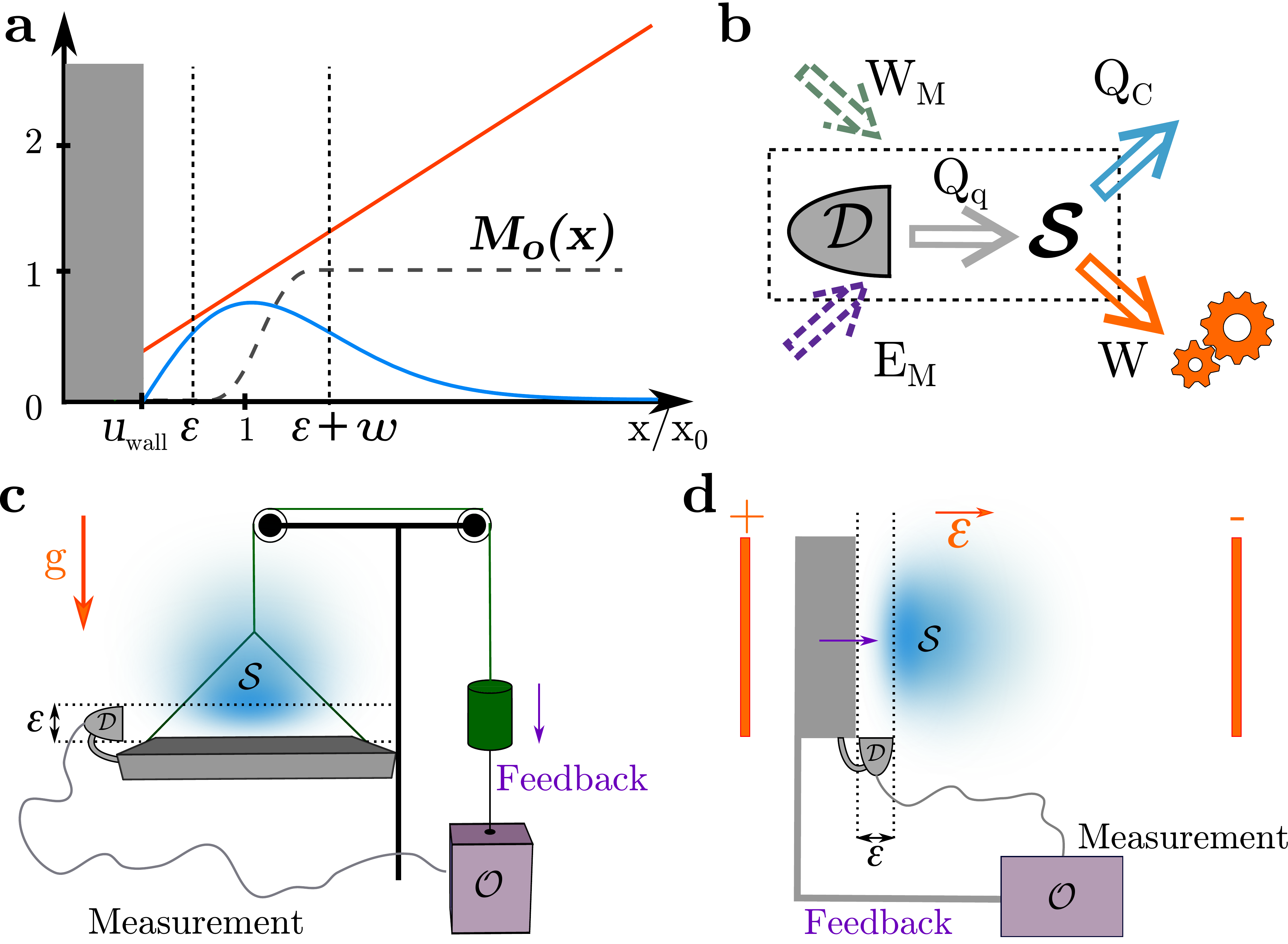}
 \caption{\textbf{a}: Situation under investigation. The particle's initial wave function (blue solid) is the ground state which is confined between the tilted potential (depicted in solid red) and the wall (in gray), located at position $x_\text{wall} = u_\text{wall} x_0$. The generalized measurement is characterized by the function $M_o(x)$ (dotted line) able to tell that, for sure, the particle is outside the region $[x_\text{wall},x_\text{wall} + \varepsilon x_0]$  when outcome $o$ is found. \textbf{b}: Energy exchanges occurring during the engine cycle: the measurement provides the quantum heat $Q_\text{q}$, which is split between  useful work $W$ and the heat $Q_\text{C}$ dissipated in the cold bath. The dotted arrows stand for the details of the energy exchanges during the measurement according to our measurement model (see SI \cite{si}): the work $W_\text{M}$ is needed to entangle $\cal S$ and the meter, and $E_\text{M}$ is the average energy provided to the meter to reset it before the next cycle. \textbf{c,d}: Two possible implementations. \textbf{c}: The elevator. An atom $\cal S$ is on a platform and experiences gravitational acceleration $g$. The detector $\cal D$ checks every cycle if the atom is within a distance $\varepsilon$ from the platform and sends the outcome to the elevator operator $\cal O$ (lift attendant) who shifts the elevator to the ``next floor'' of height $\varepsilon x_0$ for free if the outcome is $o$. \textbf{d}: The single electron battery. The negatively charged particle experiences an electric field of intensity $\cal E$ between two electrodes. The wall is a piece of neutral insulator that can be moved depending on the outcomes of $\cal D$. The electron successfully moved distance $L$ between the electrodes charges the battery with energy $e {\cal E} L$.
 }\label{f1}
 \end{figure}
 
{\it Setup.---} We will now make a quantum measurement do useful work by having a particle climb a tilted potential. The setup is the following (see Fig.~1): a particle is described by a pure state $\ket{\psi}$ in a potential $\hat V = V(\hat x) = V_0 \hat x /\xi + V_\text{wall}(\hat x)$. The term $V_\text{wall}(x)$ corresponds to a barrier of infinite height preventing the particle from reaching the positions $x<x_\text{wall}$. The time-independent Schr\"odinger equation for a particle with mass $m$ and energy $E$,
$-(\hbar^2/2m)\partial_x^2\phi(x) + V(x)\phi(x) = E\phi(x)$, can be rewritten for $x>x_\text{wall}$ as
\bb
\phi''(x) +\left(\dfrac{2mE}{\hbar^2}-\dfrac{x}{x_0^3}\right)\phi(x) = 0\label{EdSn},
\ee
with the characteristic length $x_0 = (\hbar^2\xi/2mV_0)^{1/3}$, together with the boundary condition $\phi(x_\text{wall})=0$, where $'$ denotes a spatial derivative.

The eigenstates of the Hamiltonian can therefore be expressed in term of the Airy function $\Ai(x)$ \cite{Gea99} and its zeros $\{a_l\}_{l\geq 1}$ with $a_l <0$ and $a_{l+1} <a_l$:
\be
\phi_n(x) = 
\dfrac{1}{\sqrt{x_0}}\dfrac{\Ai[(x-x_\text{wall})/x_0+a_n]}{\Ai'(a_n)},
\label{En}
\eeq
for $x\geq x_\text{wall}$, and 0 otherwise.
The energy eigenvalues are $E_n = (\hbar^2/2mx_0^2)\vert a_n\vert + (\hbar^2/mx_0^2)(x_\text{wall}/x_0)$.

Let us start the system so the particle is in the ground state $\phi_1(x)$ and the wall is at position $x=0$.  An ideal position measurement of the particle is in fact impossible, because it would require an infinite amount of energy.  Let us therefore consider another kind of position measurement, and simply determine whether the particle is within some distance $\varepsilon$ of the wall, or not.  Even this ``yes-no'' question introduces discontinuities in the wavefunction and is also too costly.  We therefore adopt a minimal model, and consider two possible outcomes of a generalized measurement, each associated with Kraus operators $M_o$ and $M_i$, where the labels $i,o$ denote that particle is found inside or outside the region $[0, \varepsilon x_0]$ from the wall.  We smooth the abrupt transition with an interpolating region from $\varepsilon x_0$ to $(\varepsilon + w)x_0$.  Let us choose $M_o$ to be
\be
M_o = \begin{cases} 0, & x/ x_0<\varepsilon, \\
\sin[\pi (x/x_0 - \varepsilon)/2w], & \varepsilon < x/ x_0< \varepsilon + w , \\
1, & x/ x_0 > \varepsilon + w .
\end{cases} \label{mo}
\eeq
$M_i^2+M_o^2=1$ for all space (let us choose $M_i$ also real) because $M_o, M_i$ are Kraus operators \cite{wisemanbook,jacobs2014quantum}, so we must then have $M_i$  decreasing from 1 at the wall as a cosine function down to 0.
Regardless of the specific form for $M_{o,i}$, quantum mechanics dictates that the probability of finding result $i, o$ is given by $P_{i,o} = \la \phi_1 | M_{i,o}^2 | \phi_1\ra = \int dx M_{i,o}^2(x) \phi_1^2(x)$, with a conditional post-measurement state given by $|\phi_{\alpha} \ra= M_\alpha |\phi\ra/\sqrt{P_\alpha}$, $\alpha = i,o$.
 
 {\it Engine cycle.---}  The three-stroke engine cycle can now be described.  
 The engine consists of the system (a single particle), a detector, and a controller to either move the wall's position or keep it in place. The object of the engine is to convert energy given by the measurement process into useful work.
\begin{enumerate}
\item A measurement of the particle's position occurs, resulting in the stochastic result $i$ or $o$ with probabilities $P_{i}, P_o$. Generally, the new (disturbed) state of the particle is no longer in its ground state and therefore has a greater internal energy, regardless of which outcome occurs. The energy gained by the particle during this step must be provided by the measurement because total energy is conserved.  We refer to the average energy gain over both outcomes, $Q_\text{q}\geq 0$, as ``quantum heat'' because of its stochastic nature.
\item If outcome $i$ was found (particle is close to the wall), then the engine controller does nothing. If outcome $o$ is found (the particle must be a distance larger than $\varepsilon$ from the wall), then the controller suddenly moves \footnote{The wall displacement is treated as a quench during which the particle wave-function does not have time to spread towards the wall.} the wall to the right of a distance $x_\text{M} = \varepsilon x_0$. This costs no work in principle because the wavefunction's value is 0.  Further, it has been shown that motion of the wall through a region of zero wavefunction makes no change to the rest \cite{mousavi2017strong}.
\item Whatever the outcome, we let the particle relax in contact with a bath of temperature $T_\text{R}$ very low with respect to $T^\ast = (E_2-E_1)/\kB\simeq 1.75 \hbar^2/2mx_0^2 k_\text{B}$ \footnote{This step can be replaced with a coherent energy extraction step with a system Hamiltonian mapping the post-measurement states to the ground state(s), so no thermal                                                                                                                                                                                                                                                                                                                                                                                                                                                                                                                                                                                                                                                                                                                       bath is required.  The required system unitaries may be included as part of the measurement operator, so that no energy is wasted on the cold bath.  However, this replacement may be impractical.}. For a relaxation time long enough, the particle is in its ground state, possibly with an advanced wall.  Note that if the wall has been advanced (outcome $o$) the new ground state has an energy increased by an amount $W = (\hbar^2/2mx_0^2) \varepsilon$: this corresponds to work extracted during the cycle from the particle's potential energy.
\end{enumerate}
 We note that there is a conditional 
 step in the above cycle, and the engine is instructed to do different things depending whether outcome $i$ or $o$ is found.  If we consider this system to be an isolated thermodynamic system of the same type of Szilard, see e.g. \cite{PhysRevLett.106.070401}
with the observer acting as a quantum Maxwell demon, the demon resets its memory in a bath of temperature $T_\text{D}$.  Although we are not extracting work from a thermal bath, but rather from the quantum measurement process, the use of finite resources used cyclically still requires erasure of memory. However, the erasure cost is $- k_\text{B}T_\text{D}\sum_{\alpha=i,o} P_\alpha \log P_\alpha$ which can be set much smaller than $W$ for sufficiently low $T_\text{D}$ (in particular for $T_\text{D} = T_\text{R}$). 

{\it Results.---}
We now analyze the engine's performance. The engine cycle is stochastic, so it is possible that from run to run, a large amount of work may be done (i.e. a long sequence of $o$ results).  However, we will consider the average performance of the engine.   The engine cycle is constructed so that the system always begins in the ground state, and therefore the average work per cycle is given by the work in steps 2 and 3, times the probability of $o$,
\be
W = \varepsilon \frac{\hbar^2}{2 m x_0^2} \int dx M_o^2(x) \phi_1^2(x). \label{work}
\eeq
The average amount of energy given by the measurement apparatus to the system (per cycle) is given by $Q_\text{q} = \sum_\alpha P_\alpha \la \phi_\alpha | H |\phi_\alpha\ra - \la \phi_1 | H |\phi_1\ra$, or
\be
Q_\text{q} = -\dfrac{\hbar^2}{2m}\int dx  \left(\sum_{\alpha} M_{\alpha}(x) M_{\alpha}''(x)\right)\phi_1^2(x),\label{heat}
\eeq
 where we have assumed $M_{o,i}$ are diagonal in the position basis, causing the potential energy term to drop out.  The conversion efficiency is defined as
 \be
 \eta = \frac{W}{Q_\text{q}}.
 \eeq
For the specific choice of the Kraus operators in Eq.~(\ref{mo}), the quantum heat takes on the simple form, ${\tilde Q}_\text{q} =  \left(\frac{\pi}{2w}\right)^2 \int_{\varepsilon x_0}^{(\varepsilon+w)x_0} dx\, \phi_1^2(x)$, where the tilde symbol denotes the work or heat divided by $\hbar^2/(2 m x_0^2)$.

\begin{figure}[t]
\begin{center}
\includegraphics[width=0.49\textwidth]{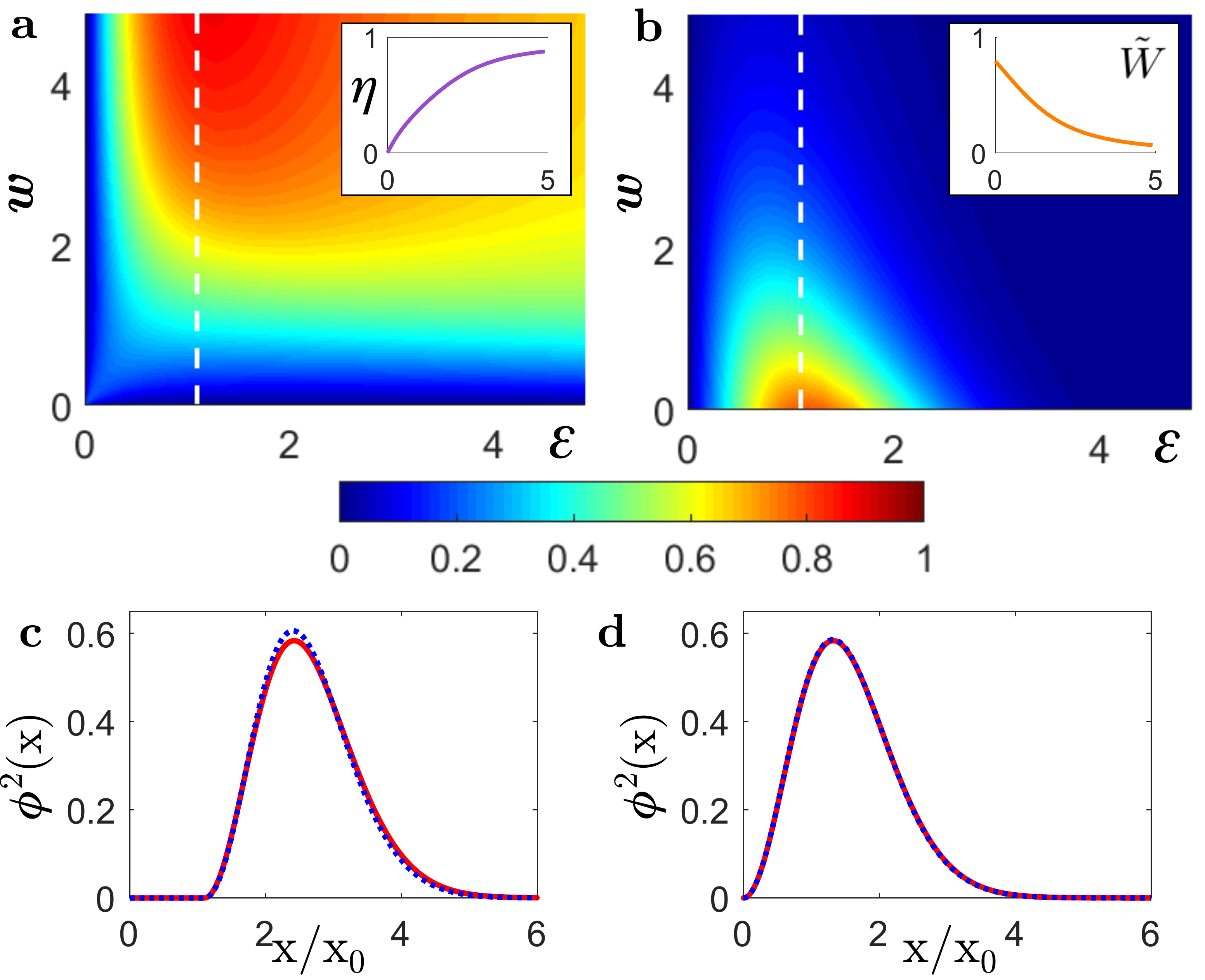}
\end{center}
\caption{\textbf{a,b}: Efficiency $\eta$ (\textbf{a}) and average work extracted per cycle $\tilde W$ (\textbf{b}) as a function of $\varepsilon$ and $w$. Insets: Section of the efficiency (\textbf{a}) and the work (\textbf{b}) along the dashed white line corresponding to $\varepsilon=\varepsilon^\ast$. \textbf{c,d}: Comparison of the post-measurement state corresponding to result $o$, panel {\bf c} ($i$, panel {\bf d}), with $\varepsilon = \varepsilon^\ast$ and $w\rightarrow \infty$ with the new ground state wavefunctions of the linear trap potential displaced by an amount $\varepsilon^\ast x_0$ (original ground state wavefunction).}\label{FigEta}\label{FigComp}
\end{figure}

 It may naively be thought that this engine is most efficient when the variables $\varepsilon, w$ are very small, such that the outcome $o$ becomes much more frequent than the outcome $i$, a kind of ``Zeno limit engine'' as in \cite{elouard2017extracting}.  In fact, this is false.  Expanding both work and heat in the limit just mentioned, we find that ${\tilde W} = \varepsilon (1 + {\cal O}(\varepsilon, w)^3)$, and ${\tilde Q_\text{q}} = (\pi/2)^2 \left( \varepsilon^2/w + \varepsilon + w/3\right)$.  Consequently, the quantum heat diverges as $w \rightarrow 0$, reflecting the problematic nature of the strict ``yes-no'' question mentioned in the setup. We can optimize the efficiency by setting $w = \zeta \varepsilon$, so that the efficiency maximizes at $\eta_{max} \approx 0.188$ when $\zeta = 1/\sqrt{3}$, even when the work limits to 0. We can do much better for the efficiency and work by choosing larger values of $\varepsilon, w$, as shown in Fig.~2.  In the plots shown there, we see that best work performance (${\tilde W} = 0.80$) is given for $\varepsilon \approx  1.18$, and for $w=0$.  However, the efficiency vanishes here since the quantum heat diverges when $w \rightarrow 0$.  We also consider the power of the engine, ${\cal P} = W/\tau$, where $\tau = P_o \tau_o + P_i \tau_i$ is the average time of the engine cycle, incorporating the time to move the wall and the thermalization time for outcome $o$ and $i$.  Let us define the mean velocity of the particle across the potential as $v = P_o \varepsilon x_0/\tau$.
Provided the relaxation in step 3. is fast enough, the cycle duration is typically proportional to $\varepsilon$ such that $v$ is finite as $\varepsilon \rightarrow 0$.  We can express the power as ${\cal P}  =(\hbar^2/2m x_0^2) (v/x_0) = v V_0/\xi$, or the particle's velocity times the force acting on it, allowing finite power at finite efficiency in the Zeno limit where no memory or feedback is required because every outcome is almost certainly $o$ \footnote{Supposing the height of the building is $L$, there are $N = L/(x_0 \varepsilon)$ floors that must be crossed.  At the ideal efficiency of $\eta_{max}$ the probability of getting to the top floor with no stops is given by $\approx \exp[-(0.327/N^2) (L/x_0)^3]$.  Consequently, for buildings of fixed height, we can always make the number of floors $N$ sufficiently large to almost always have deterministic operation.}, 
 \footnote{We have also investigated the engine performance with Kraus operators $M_{i,o}$ with continuous second derivatives via a smoothed step function, and found similar results as the ones reported here.}.

{\it Gradual measurement limit.}---
We notice in the Fig.~\ref{FigEta}(a), that we get more efficient engines for large values of $w$ and moderate values of $\varepsilon$.  Large values of $w$ correspond to measurements that have a very slow turn-on outside the ``window'' region of $[0, x_0\varepsilon]$.
Define the asymptotic efficiency, $\eta_{asyp} = \lim_{w\rightarrow \infty} \eta$ to find \footnote{Mathematically, it is not at all obvious that this efficiency must be less than 1.  Indeed, if we replace $\phi^2_1(x)$ by other normalized distributions, this will not hold, so it is special for the Airy wavefunction.}  
 \be
 \eta_{asyp} = \varepsilon\, \frac{
 \int_{\varepsilon x_0}^\infty dx \, (x/x_0 - \varepsilon)^2 \phi^2_1(x)
 }{ \int_{\varepsilon x_0}^\infty dx \,  \phi^2_1(x) }. \label{asyp}
 \eeq
Incredibly, the efficiency of the measurement approaches 1 at $\varepsilon =\varepsilon^\ast \approx 1.100$, corresponding to a maximal efficiency of $\eta_{max} \approx 0.998$.
We can understand more deeply why this optimal value of $\varepsilon^\ast$ corresponds to maximal efficiency by plotting the (normalized) post-measurement state $\phi_o(x) \propto (x-\varepsilon^\ast)\Ai(x+a_1)$, 
and comparing it to the new ground state of the (displaced) trap potential ($\propto \Ai(x+a_1 - \varepsilon x_0)$), shown in Fig.~2{\bf c}; they are nearly identical. Similarly, the comparison between $M_i \phi_1(x)$ with the original ground state look nearly the same, Fig.~2{\bf d}.  Consequently, either moving the wall by $\varepsilon x_0$ or leaving it in place almost perfectly realizes the next phase of the engine cycle
\footnote{It is also interesting to see how this limit is approached (since the work is strictly 0 in the $w \rightarrow \infty$ limit).  Fixing $\varepsilon = \varepsilon^\ast$, and varying $w$, we find the efficiency $\eta$ exceeds 0.9 when $w = 5 x_0$ and the scaled work is ${\tilde W
} = 0.065$ (corresponding to a 5.9\% success probability), and $\eta$ exceeds 0.99 at $w=17 x_0$, where the scaled work is ${\tilde W} = 0.0061$ (corresponding to a success probability of 0.55\%).}.

 
{\it Implementations.---} We implement two different variations of the engine shown in Fig.~1.  The first is a single atom elevator: a gravitational potential of $V(x) = m g x$ acts on the atom, resulting in the characteristic length of $x_0 = (\hbar^2/2 m^2 g)^{1/3} \approx 6 \mathrm{\mu m}/{\tilde m}^{2/3}$ for an atom near the surface of the earth of relative atomic mass ${\tilde m}$. The temperature needed to cool to the ground state is $T^\ast = {\tilde m}^{1/3} 12\mathrm{nK}$, so for e.g. a Rb atom, we have $x_{0,Rb} \approx 300 nm$ and require $T^\ast_{Rb} \approx 50 nK$, which is quite possible to realize in cold atom experiments. Alternatively, we can consider one ultra-cold neutron above a neutron mirror, which is the setup of recent gravity-resonance spectroscopy experiments \citep{Jenke11,Cronenber15}.
In the sketch of Fig.~1{\bf a}, the elevator has a platform that has a counter balanced weight over the pulley.  Since the net force is zero, the elevator can be raised to the ``next floor'' by the elevator operator with no work done, so long as the movement only occurs when the atom has no amplitude to be near the platform.   

In our second example shown in Fig.~1{\bf b}, we consider a parallel plate capacitor that is being charged, one electron at a time (a battery).  We consider a potential difference of 1$V$ across a 1$cm$ gap.  This gives a characteristic length scale of $x_0 = (\hbar^2/(2 m e^2 {\cal E}))^{1/3} \approx 72 nm$, where $\cal E$ is the electric field between the plates. The required thermalization temperature is only $T^\ast \approx 0.15 K$ because the electron is so light.  An insulating, uncharged plate with negligible susceptibility can be moved through the electric field without any work done.  The plate stops the electron from accelerating back to the positively charged plate.  A measurement of the electron's position away from the plate allows the controller to advance the position of the plate to bring the electron to the other side of the capacitor, charging the battery. 

In the SI \cite{si}, we present a model of the measurement, implemented by a spin-1/2 meter impulsively interacting with the particle, in order to track the energy exchange.  Letting the spin begin with energy $E_o - E_i$ the difference of the energies of the states corresponding to outcomes $o$ and $i$, during the interaction, an amount of work $W_\text{M} = E_i - E_1 \approx 0$  is performed on the joint spin-particle system. The average energy given away by the spin is $E_\text{M} = P_o (E_o-E_i)$.  These energies provide the ``fuel'' for the quantum measurement engine, $Q_\text{q} = E_\text{M}+ W_\text{M}$, and must be replenished for the engine to continue working, see Fig.~1{\bf b}: as dictated by the WAY theorem, the measurement is not repeatable if the meter is not externally powered.

{\it Conclusions.---} We have constructed an explicit quantum engine that converts energy from quantum measurement to do useful work on the system.  This process requires feedback in general.  We stress that a simple transfer of energy is not sufficient to make a working engine.  The energy must be transferred in such a way that it can be efficiently extracted.  To this end, our three stroke engine is near optimal because one outcome produces nearly the correct ground state of the system in the next cycle, while the other outcomes leaves the state nearly the same as before.  The ability to advance our wall with no work expended allows efficient conversion of kinetic to potential energy to make the particle do work against an opposing force, provided by the measurement process.  In spite of the stochastic nature of the measurement process, we are able to attain efficiencies approaching unity.  This result clearly illustrates the differences with quantum thermodynamic systems. 

{\it Acknowledgments.---} This work was supported by the US Department of Energy grant No. DE-SC0017890. We thank Chapman University and the Institute for Quantum Studies for hospitality during this project. We thank Rafael Sanchez and Yunjin Choi for helpful discussions.

\bibliography{Biblio}
\pagebreak
\section{Supplementary Information}
Let us discuss a possible implementation of the measurement process.  We consider the interaction between the particle in the potential well and a spin 1/2 particle with the impulsive interaction Hamiltonian 
\be
H_{int} = \hbar f({\hat x)} \sigma_y \delta(t),
\eeq
where the spin is prepared in the state $|0\ra$, and $f$ is some function of the particle's coordinate.  After the interaction takes place, the spin is measured in the basis $\{\ket{0},\ket{1}\}$. Starting in the separable state of the spin and particle given by $\ket{0}\ket{\phi}$, the interaction entangles the spin with the particle's position, leaving the state
\be
\ket{\Psi} = \ket{0} \cos[f({\hat x})]|\phi\ra
+\ket{1} \sin[f({\hat x})]|\phi\ra.
\eeq 
It is now clear that the Kraus operators are $M_{0,1} = \{ \cos[f({\hat x})], \sin[f({\hat x})] \}$.  These can reproduce the investigated set (\ref{mo}) by choosing $f(x)$ to be 0 for $(x -x_\text{wall})/x_0 < \varepsilon$, $\pi (x/x_0-\varepsilon)/2 w$ for $\varepsilon < x/x_0 < \varepsilon + w$, and $\pi/2$ for $x/x_0 > \varepsilon + w$, {\it i.e.} a linear ramp from 0 to $\pi/2$ in the transition region.
The linear ramp can be implemented, {\it e.g.} optically, replacing the spin by the polarization degree of freedom, and by having the polarization of a photon rotated by $90^\circ$ if it interacts with the particle.  Overlaying a transmission mask that changes from transmissive to reflective will then 
implement the interaction Hamiltonian; non-demolition measurements of the photon's polarization can then be made to realize the generalized measurement. This model allows us to track the energy exchanges occurring during the measurement. Assuming a spin with energy splitting given by $\hbar\omega_\text{M} = E_o - E_i$ in the $|0\ra, |1\ra$ basis, where $E_\alpha$ is the energy of state $\phi_\alpha$ for $\alpha = i,o$, we find that during the interaction, an amount of work $W_\text{M} = E_i - E_1$  is performed on the joint spin-particle system. This work is negligible in the case of an optimal measurement scheme fulfilling $\ket{\phi_i}\simeq \ket{\phi}$. During the interaction, the system energy increases by an average amount $\Delta U_1 = P_o E_o + P_i E_i - E_1$. Then, when the spin meter is measured, a quantum of energy $\hbar\omega_\text{M}$ is randomly relocalized either in the spin (outcome $i$) or in the particle (outcome $o$). This corresponds to a change of the particle's energy by $\Delta U_2^{(\alpha)} = E_\alpha - (P_o E_o +P_i E_i)$ 
for outcome $\alpha = i,o$. Considering the total change in the system energy in the two-step process, $\Delta U_1 + \Delta U_2^{(\alpha)}$, and averaging over the two measurement outcomes recovers the average quantum heat $Q_\text{q}$ of Eq.~\eqref{heat}. We note that if outcome $o$ is found, the spin is left in its ground state $\ket{1}$ and must be reset to the excited state $\ket{0}$ before the next measurement cycle if we reuse that spin. This process costs a amount of energy $\hbar \omega_\text{M}$ which together with the interaction work $W_\text{M}$  compensates the quantum heat $Q_\text{q}$ provided by the measurement process, and is therefore not repeatable in isolation as dictated by the WAY theorem. 

\end{document}